\documentclass[conference]{IEEEtran}
\IEEEoverridecommandlockouts
\usepackage{amsmath,amssymb}
\usepackage{graphicx}
\usepackage{booktabs}
\usepackage{cite}
\usepackage{url}
\usepackage{hyperref}
\hypersetup{colorlinks=true, urlcolor=blue, linkcolor=black, citecolor=black}
\makeatother
\makeatletter
\g@addto@macro\normalsize{%
  \setlength{\abovedisplayskip}{4pt}%
  \setlength{\belowdisplayskip}{4pt}%
  \setlength{\abovedisplayshortskip}{4pt}%
  \setlength{\belowdisplayshortskip}{4pt}%
}
\makeatother

\begin{document}

\title{Grid-Oriented Probabilistic Load Forecasting\\for AI Data Centers}

\author{\IEEEauthorblockN{Qingyi Yao\textsuperscript{1}, Ding Lin\textsuperscript{1},
Tianqiao Zhao\textsuperscript{2}, Meng Yue\textsuperscript{3}, and Jianhui Wang\textsuperscript{1}}
\IEEEauthorblockA{\textsuperscript{1}Southern Methodist University, Dallas, TX, USA\\
\textsuperscript{2}The University of Texas at Arlington, Arlington, TX, USA\\
\textsuperscript{3}Brookhaven National Laboratory, Upton, NY, USA}
}

\maketitle

\begin{abstract}
Probabilistic load forecasts for AI data centers need to represent both uncertainty under
changing workloads and temporal dependence to support grid operations. This paper develops an
open, grid-oriented benchmark using measured GPU power from a 448-GPU AI cluster. Forecasts are
evaluated through four stages: probabilistic prediction, online uncertainty calibration, temporal
scenario generation, and horizon-wide reserve sizing. Following a change in cluster utilization,
a quantile LightGBM forecast achieves only 63.4\% coverage for a nominal 90\% interval, while
online regime-conditioned calibration increases coverage to 87.3\%. Even with well-calibrated
marginal forecasts, independent sampling produces unrealistic ramp distributions and excessive
reserve procurement. Reconstructing temporal dependence from historical forecast errors reduces
the discrepancy in ramp statistics and lowers realized reserve cost by 7.4--10.6\% across three
forecasting tasks. These findings show that marginal forecast quality alone is insufficient for
operational decisions and motivate joint evaluation of calibration, temporal structure, and
decision outcomes.
\end{abstract}

\vspace{-0.2cm}
\section{Introduction}

Electricity demand from data centers is growing quickly, and artificial intelligence (AI)
training and inference on graphics processing units (GPUs) account
for a rising share of it \cite{shehabi2024lbnl,iea2025energyai}. For utilities and system
operators the difficulty lies less in the expected level of this load than in its uncertainty.
They need to know how fast a facility ramps, how long it stays near its peak, and how often it
departs from a schedule. Recent industry studies describe data-center demand through multiple growth
scenarios \cite{epri2026powering} and identify the uncertainty of large-load forecasts as a risk
for resource and transmission planning \cite{esig2025largeloads}.

Existing studies on data-center load forecasting have employed statistical models, recurrent, convolutional and attention-based networks, probabilistic methods, and physics-informed hybrids \cite{shekhar2026review}. However, many studies emphasize point forecast accuracy and rely on proprietary or synthetic datasets. Their results are therefore difficult to compare, and conventional accuracy metrics alone do not establish whether a forecast is suitable for grid studies.

Forecast accuracy alone does not establish decision value in power system applications
\cite{habib2026bridging}.  For grid studies, two aspects of probabilistic load forecasts are particularly important. The first is reliable uncertainty calibration under changing workload conditions, ensuring that prediction intervals maintain their nominal coverage. The second is realistic temporal dependence across the forecast horizon, since load ramps, peaks, and high-load durations directly affect ramping and reserve requirements. A forecast may achieve a low continuous ranked probability score (CRPS) at each step while failing to maintain reliable coverage or reproduce realistic temporal dependence.

Evaluating these properties requires assessing both marginal forecast calibration and the temporal structure of generated load scenarios, while meaningful comparisons require accessible data. This paper therefore develops an open benchmark using GPU telemetry and scheduler logs from the MIT Supercloud TX-Gaia cluster \cite{samsi2021supercloud} and evaluates forecasts stage by stage, from probabilistic forecasting to reserve sizing.
The main contributions are summarized as follows:
\begin{enumerate}
\item An open benchmark for grid-oriented probabilistic forecasting of AI-cluster GPU
power. Using measured GPU telemetry and scheduler logs, we define three forecasting tasks from
intra-hour to day-ahead horizons and provide reproducible data processing and evaluation that
score ramp, peak and reserve outcomes alongside conventional probabilistic metrics.
\item A four-stage evaluation framework that links probabilistic forecasts to reserve
decisions. The framework evaluates probabilistic prediction, online calibration,
temporal scenario generation, and horizon-wide reserve sizing using stage-specific metrics. Its
modular structure allows different forecasters and scenario generators to be evaluated without
modifying the remaining stages.
\item An empirical assessment of forecast calibration, temporal dependence, and their
implications for reserve decisions. Online calibration improves coverage from 63.4\% to 87.3\% for an under-calibrated forecaster but offers limited benefit to one already near nominal coverage. Despite near-nominal marginal coverage, independent scenario
sampling distorts load ramps and leads to excessive reserve procurement. Reconstructing temporal
dependence from past forecast errors while preserving the conditional marginal distributions
lowers realized reserve cost by 7.4--10.6\% across three forecasting tasks.
\end{enumerate}
The processing, baseline and evaluation code is available at
\url{https://github.com/dinglin613/Data_center_forecasting}.

\section{Data and Benchmark Tasks}

\subsection{Cluster Telemetry}
TX-Gaia is a production AI cluster with 224 nodes, each with two NVIDIA V100 GPUs
\cite{samsi2021supercloud}.  The public MIT Supercloud dataset contains power measurements for each GPU assigned to a job at 100-ms resolution, together with Slurm scheduler logs recording job submission, start and end times, and allocated resources.  The records span 5~Feb. to
1~Oct.~2021.

\subsection{Target Series}
For each job and node, we average the power of each GPU over 1-min intervals and sum over the
GPUs. Summing over all jobs then gives the measured GPU power of the cluster. GPUs not assigned to
any job draw idle power that the telemetry does not record. We add an assumed 40~W for each such
GPU, which sets a floor of about 18~kW when the cluster is empty. The target is therefore cluster
GPU IT power, which excludes CPU, cooling and facility losses.

The GPU timestamps record US Eastern wall-clock time. The scheduler records UTC. Matching
job start and end times reveals a 5-h offset before 14~Mar.~2021 and a 4-h offset after. We use
these offsets to convert the GPU timestamps to UTC. After this correction, the ratio of GPUs with
telemetry to GPUs allocated by the scheduler has a median of 1.00 and is at least 0.86 in 90\% of
minutes.
{
\setlength{\abovecaptionskip}{0pt}
\setlength{\belowcaptionskip}{-15pt}
\begin{figure}[!t]
  \centering
  \includegraphics{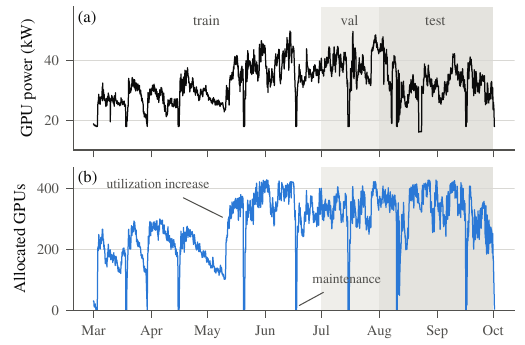}
  \caption{MIT Supercloud TX-Gaia cluster, March--September 2021, hourly means.
  (a) GPU IT power, including an assumed 40~W idle draw for unallocated GPUs.
  (b) GPUs allocated by the scheduler. Shading marks the training, validation and test periods.}
  \label{fig:data}
\end{figure}}

Fig.~\ref{fig:data} shows the hourly series and the allocated GPUs. The series has a mean of
31.2~kW, a 1-min maximum of 66.1~kW, and a largest 5-min change of 28.3~kW. Jobs are allocated
groups of GPUs, and jobs longer than one day account for 70\% of GPU-hours, so each start or
completion shifts the load for as long as the job runs. Changes in allocated GPUs explain only
31\% of the variance of 5-min power changes, and the rest comes from power changes within running
jobs. The cluster empties for about a day every two to five weeks, consistent with scheduled
maintenance. In mid-May the weekly mean of allocated GPUs rises from about 250 to about 350. The
training data thus span both utilization levels, and the test period lies entirely at the higher
level.

\subsection{Tasks, Splits and Workload Regimes}
Three tasks span intra-hour to day-ahead horizons. They pair 5-min resolution with a 1-h horizon
($H=12$), 15-min resolution with a 4-h horizon ($H=16$), and 1-h resolution with a 24-h horizon
($H=24$), where $H$ is the number of forecast steps. A forecast is issued at every step. The splits
are chronological, with training from 1~Mar. to 30~Jun., validation in July, and test from 1~Aug.
to 30~Sep.~2021. An origin belongs to a split only if its whole target window lies in that split,
so no target is shared across splits. The test period contains 17{,}556, 5{,}840 and 1{,}440
origins for the three tasks.

Each origin is assigned one of four workload regimes from information available at that time.
Origins with fewer than 25\% of the GPUs allocated are labeled \emph{low}. These fall mostly in
maintenance windows such as the one marked in Fig.~\ref{fig:data}. The other origins are labeled
\emph{calm}, \emph{active} or \emph{volatile} by terciles of the mean absolute power change over
the preceding hour, with thresholds fitted on the training period. Among these origins, any with
more than 448 GPUs requested in the queue is labeled volatile regardless of its recent power
changes. In the test period the low, calm, active and volatile regimes cover about 3\%, 17\%,
30\% and 50\% of the origins.

\section{Grid-Oriented Evaluation Framework}

Fig.~\ref{fig:framework} shows the framework. Each stage receives the output of the previous one,
passes its own output on, and is scored with its own metrics. Any component in a stage can be
replaced, so a new forecaster or scenario generator can be evaluated without changing the other
stages.
{
\setlength{\abovecaptionskip}{0pt}
\setlength{\belowcaptionskip}{-15pt}
\begin{figure*}[!t]
  \centering
  \includegraphics{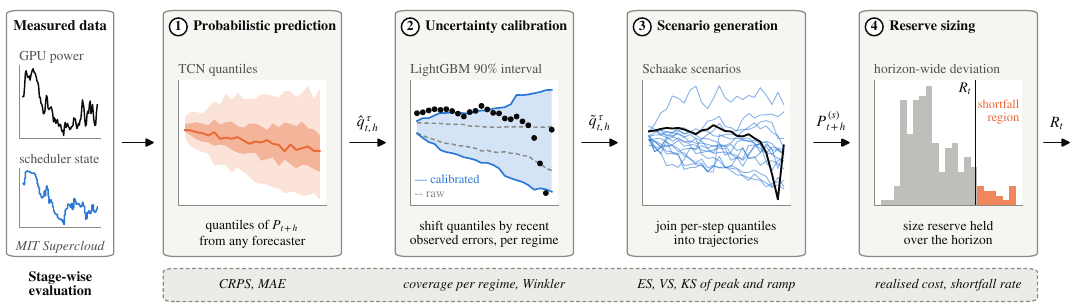}
  \caption{Grid-oriented evaluation framework. Mini-plots show each stage's output for the test
  origin 2021-08-02 11:00 UTC (1-h resolution, 24-h horizon). Each stage passes its output to the
  next and is scored with its own metrics (ES: energy score, VS: variogram score,
  KS: Kolmogorov--Smirnov distance).}
  \label{fig:framework}
\end{figure*}}

\subsection{Stage 1: Probabilistic Prediction}
At origin $t$, a forecaster returns the quantiles $\hat q^{\,\tau}_{t,h}$ of the power
$P_{t+h}$ for $h=1,\dots,H$ and the nine levels
$\tau\in\{0.025,0.05,0.1,0.25,0.5,0.75,0.9,0.95,0.975\}$. The pinball loss of a quantile
forecast is
\begin{equation}
\rho_\tau(u)=u\,\bigl(\tau-\mathbb 1\{u<0\}\bigr),\qquad u=P_{t+h}-\hat q^{\,\tau}_{t,h}.
\end{equation}
We evaluate four forecasters. \emph{Persistence} repeats the latest observed value and adds the
empirical quantiles of its training errors at each horizon. \emph{Seasonal naive} uses the value at the
same time on the previous day and forms its quantiles the same way. \emph{LightGBM}
\cite{ke2017lightgbm} fits one gradient-boosted model per quantile level, each minimizing the
pinball loss at that level, to predict the change in power from the forecast origin, with the
forecast horizon included as an input. Input features include recent power changes, rolling means
and spreads, allocated and queued GPU fractions, power per allocated GPU, and hour of day and day
of week. The temporal convolutional network (\emph{TCN}) \cite{bai2018tcn} processes the same features
over a 24-h lookback using seven residual blocks of dilated causal convolutions (kernel size 3, 64
channels, and dilations from 1 to 64). It jointly predicts all $H\times9$ quantiles of the power
change from the forecast origin. The TCN is trained on the mean pinball loss over all levels and
horizons with early stopping on validation loss, and its results are averaged over three random
seeds.

\subsection{Stage 2: Online Uncertainty Calibration}
A forecaster may become miscalibrated as workload conditions change. Online conformal methods use
recent forecast errors to improve calibration under changing conditions. The
calibrator shifts each quantile by an empirical quantile of these errors. With the score
$s^{\tau}_{j,h}=P_{j+h}-\hat q^{\,\tau}_{j,h}$, the calibrated quantile is
\begin{equation}
\tilde q^{\,\tau}_{t,h}=\hat q^{\,\tau}_{t,h}+\mathrm Q_\tau\bigl\{s^{\tau}_{j,h}:j\in
\mathcal C_{t,h}\bigr\},
\end{equation}
\begin{equation}
\mathcal C_{t,h}=\{\,j:\ t-h-W\le j<t-h,\ r_j=r_t\,\},
\end{equation}
where $\mathrm Q_\tau$ is the empirical $\tau$-quantile, $W$ is the number of origins in 14 days
and $r_t$ is the workload regime. The condition $j<t-h$ admits only origins whose targets are
observed before $t$, so the correction uses no future information. If fewer than 200 scores in the
window share the current regime, all scores in the window are used. The calibrated quantiles are
then sorted, so they do not cross. At the first test origin, the window contains the last 14 days
of validation data. Corrections are refreshed hourly. Stage-2 metrics are
the empirical coverage of the central 90\% interval, overall and per regime, and the Winkler
interval score.

\subsection{Stage 3: Temporal Scenario Generation}
Quantiles at each step do not define a joint distribution over the horizon. Ramps, high-load
duration and energy depend on how errors at consecutive steps move together. A scenario generator
turns the input quantiles, raw or calibrated, into $S=200$ trajectories $P^{(s)}_{t+1:t+H}$. Let
$F^{-1}_{t,h}$ be the quantile function interpolated from these quantiles. We compare four
generators. \emph{Independent} sampling draws each step separately from $F^{-1}_{t,h}$. The
\emph{Gaussian copula} draws correlated normal variables with a correlation
matrix estimated on the validation period. The \emph{residual bootstrap} adds to the median
forecast whole past error trajectories $e_j=P_{j+1:j+H}-\hat q^{\,0.5}_{j,1:H}$, drawn from
origins whose whole horizon has already been observed. The \emph{Schaake shuffle}
\cite{clark2004schaake} keeps the conditional marginals and borrows only the rank structure of
past errors. It draws $x_{h,(s)}=F^{-1}_{t,h}\bigl((s-\tfrac12)/S\bigr)$ at each step and selects
$S$ of these trajectories at random. At each step $h$, scenario $s$ receives
the sample whose rank equals the rank of $e_{j_s,h}$ among the selected errors.

The energy score (ES) and the variogram score (VS) of order 0.5 assess the joint predictive distribution represented by the scenarios. We
also compute grid statistics of each trajectory: peak power, maximum ramp-up and ramp-down,
energy, and the longest duration above the 90th percentile of validation power. For each
statistic we compute the probability integral transform (PIT) of the observed value under the scenario distribution, randomized
for ties. We report the Kolmogorov--Smirnov (KS) distance of these PIT values from uniform to
assess how well the scenarios reproduce each statistic.

\subsection{Stage 4: Horizon-Wide Reserve Sizing}
The scheduled power is the median of the input quantiles, $\hat P_{t+h}=q^{\,0.5}_{t,h}$, and a reserve
$R_t$ covers positive deviations held over the whole horizon. With reserve cost $c_R$ per kW and
shortfall cost $c_S$ per kW, the realized cost is
\begin{equation}
c_R R_t + c_S\bigl(D_t-R_t\bigr)^{+},\qquad D_t=\max_{h}\bigl(P_{t+h}-\hat P_{t+h}\bigr).
\end{equation}
The expected cost is minimized by the $(1-c_R/c_S)$-quantile of $D_t$, corresponding to the newsvendor critical fractile. We estimate the reserve from the scenarios as
$R_t=\mathrm Q_{1-c_R/c_S}\{\max_h(P^{(s)}_{t+h}-\hat P_{t+h})\}_{s=1}^S$. The maximum over the
horizon depends on the joint behavior of the steps, so this decision tests Stage 3 directly. We
set $c_S=1$ and $c_R/c_S\in\{0.05,0.1,0.2\}$. At $c_R/c_S=0.1$ the target shortfall rate is
10\%. Stage-4
metrics are the realized mean cost (in kW-equivalent units) and the shortfall rate.

\section{Results}

All results are on the test period. Stage-3 and Stage-4 results use 3000 randomly selected test
origins for the 5-min and 15-min tasks and all 1{,}440 origins for the 24-h task.

\subsection{Forecast Skill Across Horizons}
Table~\ref{tab:forecast} reports Stages 1 and 2. Seasonal naive is omitted from the table. Its
CRPS of 2.23--2.30~kW is 1.5 to 5 times that of
the TCN, because the cluster load has little daily periodicity (Fig.~\ref{fig:data}).
Persistence has the lowest mean absolute error (MAE) at 5-min
and 15-min resolution. The TCN has the lowest CRPS in all three
tasks, 0.438, 0.741 and 1.462~kW, with a standard deviation over seeds of at most 0.006~kW.
LightGBM improves on persistence in CRPS at 5-min resolution only, and at the 24-h horizon its
CRPS of 1.746~kW is higher than that of persistence.

\begin{table}[!t]
  \centering
  \caption{Forecast skill and calibration on the test period (Aug.--Sep.\ 2021)}
  \label{tab:forecast}
  \footnotesize
\begin{tabular}{@{}lrrrrr@{}}
\toprule
Forecaster & CRPS & MAE & Cov. & Min cov. & Winkler \\
 & (kW) & (kW) & (\%) & (\%) & \\
\midrule
\multicolumn{6}{@{}l}{\emph{5-min / 1-h}} \\
Persistence & 0.504 & 0.738 & 82.9 & 68.0 & 8.01 \\
LightGBM & 0.481 & 0.792 & 82.4 & 73.1 & 7.07 \\
LightGBM + OC & 0.469 & 0.783 & 87.9 & 77.2 & 6.80 \\
TCN & \textbf{0.438} & 0.747 & 88.3 & 86.8 & 6.11 \\
TCN + OC & \textbf{0.438} & 0.746 & \textbf{88.4} & 87.3 & 6.13 \\
\addlinespace
\multicolumn{6}{@{}l}{\emph{15-min / 4-h}} \\
Persistence & 0.809 & 1.286 & 83.2 & 62.6 & 12.04 \\
LightGBM & 0.823 & 1.360 & 73.9 & 59.3 & 12.07 \\
LightGBM + OC & 0.769 & 1.328 & 87.9 & 71.3 & 10.81 \\
TCN & \textbf{0.741} & 1.313 & \textbf{88.5} & 84.4 & 9.98 \\
TCN + OC & 0.751 & 1.319 & 87.9 & 83.9 & 10.15 \\
\addlinespace
\multicolumn{6}{@{}l}{\emph{1-h / 24-h}} \\
Persistence & 1.607 & 2.801 & 85.7 & 50.7 & 21.95 \\
LightGBM & 1.746 & 2.844 & 63.4 & 57.3 & 25.91 \\
LightGBM + OC & 1.551 & 2.850 & 87.3 & 82.2 & 20.69 \\
TCN & \textbf{1.462} & 2.739 & 87.6 & 84.3 & 18.59 \\
TCN + OC & 1.535 & 2.841 & \textbf{87.8} & 83.1 & 20.17 \\
\bottomrule
\end{tabular}
  \\[2pt]
  \parbox{\columnwidth}{\scriptsize OC: online regime-conditioned calibration. Cov.: empirical
  coverage of the central 90\% interval. Min cov.: lowest coverage over the four workload regimes.
  TCN rows are means of three seeds. Bold: lowest CRPS and coverage closest to 90\% in each task.}
\end{table}

\subsection{Calibration Under Changing Workload Conditions}
Without correction, LightGBM covers 82.4\%, 73.9\% and 63.4\% of outcomes with its nominal
90\% interval across the three tasks. At the 24-h horizon its mean interval width is 5.4~kW, half the
10.9~kW of the TCN. About 61\% of the training period precedes the mid-May rise in utilization, and the narrow test intervals are consistent with error spreads learned at the lower level. The deficit appears in every regime
(Fig.~\ref{fig:regime}) and is not confined to the low regime.

The online correction raises LightGBM coverage to 87.9\%, 87.9\% and 87.3\%, and the lowest
regime coverage at the 24-h horizon from 57.3\% to 82.2\%. CRPS falls from 1.746 to 1.551~kW
and the Winkler score from 25.91 to 20.69. Regime conditioning mainly evens coverage across regimes. At 5-min resolution, the calm, active and volatile regimes
reach 93.9\%, 89.5\% and 87.3\% with an unconditional window and 88.4\%, 88.4\% and 88.0\% with
the regime condition.

The TCN covers 88.3\%, 88.5\% and 87.6\% without correction, and at least 84.3\% in every
regime. Its 24-h lookback contains the current utilization level and the recent volatility, so its
quantiles are conditioned on both at the origin. Applied to the TCN, the correction changes
coverage by less than one point and widens the 24-h interval from 10.9 to 12.6~kW, which raises CRPS from 1.462 to
1.535~kW. The correction is therefore a safety net with a small cost. It recovers most of the
coverage lost by a forecaster that does not track the regime, and it adds width to one that
already does.
{
\setlength{\abovecaptionskip}{0pt}
\setlength{\belowcaptionskip}{-15pt}
\begin{figure}[!t]
  \centering
  \includegraphics{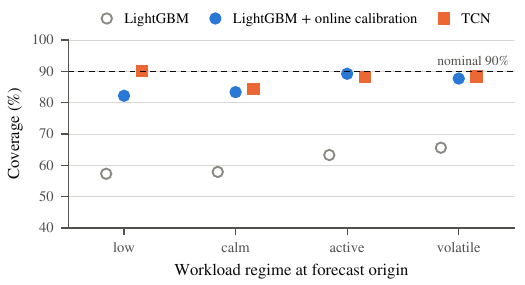}
  \caption{Empirical coverage of the central 90\% interval by workload regime at the forecast
  origin, 1-h resolution, 24-h horizon, test period. }
  \label{fig:regime}
\end{figure}}
{
\setlength{\abovecaptionskip}{0pt}
\setlength{\belowcaptionskip}{-15pt}
\begin{figure*}[!t]
  \centering
  \includegraphics{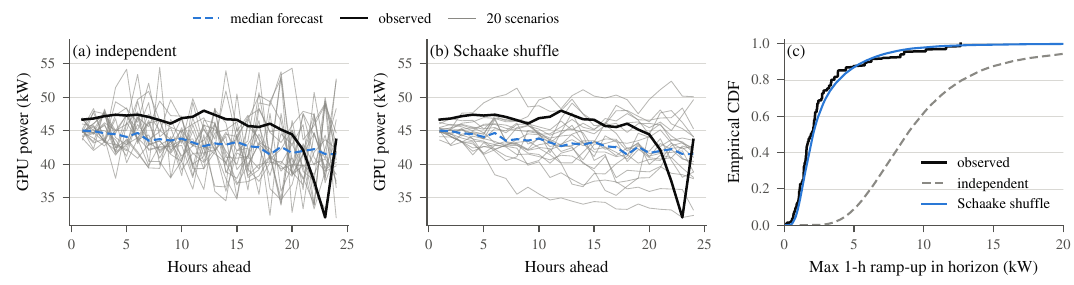}
  \caption{Scenarios from the quantile TCN, 1-h resolution, 24-h horizon.
  (a), (b) Twenty scenarios for the test origin 2021-08-02 11:00 UTC, selected as the largest
  observed 1-h ramp-up among origins outside the low regime whose observed window stays above 20~kW.
  (c) Empirical CDF of the maximum 1-h ramp-up within the horizon over all 1440 test origins.}
  \label{fig:scenarios}
\end{figure*}}

\subsection{Temporal Structure of the Scenarios}
Table~\ref{tab:scenarios} reports Stages 3 and 4 for the uncalibrated TCN quantiles. Independent sampling has
the largest ES and VS in every task, and its KS distances for the peak (0.52--0.66) and the
largest ramp-up (0.70--0.88) show that the observed statistics are concentrated at one end of
the simulated distribution. Consecutive TCN forecast errors are strongly correlated, with a mean lag-1 correlation of 0.84,
0.85 and 0.90 in the three tasks. Sampling the steps separately removes this correlation, so
simulated trajectories move up and down from step to step far more than the observed load
(Fig.~\ref{fig:scenarios}a, c).

All three generators that restore dependence reduce these distances. The Gaussian copula fixes
most of the peak distribution (KS 0.14--0.16) but leaves ramps biased (0.35--0.42). A single Gaussian dependence structure, estimated once on
July, does not reproduce the distribution of step-to-step changes. The residual bootstrap
reproduces ramps and peaks at 5-min and 15-min resolution. It adds past errors without rescaling them to
the forecast spread at each origin, and so ignores the conditional spread of the forecaster. The Schaake shuffle combines the conditional marginals of the TCN with the empirical
rank structure. It has the lowest ES and VS in all three tasks and KS distances of 0.07--0.13 for
ramp-up and 0.08--0.09 for the peak.

\begin{table}[!t]
  \centering
  \caption{Scenario quality and horizon-wide reserve, quantile TCN, test period}
  \label{tab:scenarios}
  \footnotesize
  \setlength{\tabcolsep}{4pt}
\begin{tabular}{@{}lrrrrrr@{}}
\toprule
Generator & ES & VS & KS & KS & Reserve & Short- \\
 & & & peak & ramp & cost & fall (\%) \\
\midrule
\multicolumn{7}{@{}l}{\emph{5-min / 1-h}} \\
Independent & 2.38 & 17.7 & 0.52 & 0.70 & 0.353 & 4.7 \\
Gaussian copula & \textbf{2.30} & 13.0 & 0.14 & 0.35 & 0.333 & 7.1 \\
Residual bootstrap & 2.43 & 13.7 & \textbf{0.06} & \textbf{0.07} & 0.390 & 11.0 \\
Schaake shuffle & \textbf{2.30} & \textbf{12.5} & 0.08 & \textbf{0.07} & \textbf{0.326} & \textbf{10.5} \\
\addlinespace
\multicolumn{7}{@{}l}{\emph{15-min / 4-h}} \\
Independent & 4.83 & 60.5 & 0.62 & 0.81 & 0.668 & 4.2 \\
Gaussian copula & 4.61 & 38.5 & 0.16 & 0.38 & 0.621 & 7.3 \\
Residual bootstrap & 4.75 & 37.4 & \textbf{0.07} & \textbf{0.12} & 0.714 & \textbf{10.2} \\
Schaake shuffle & \textbf{4.58} & \textbf{36.5} & 0.09 & 0.13 & \textbf{0.607} & 9.3 \\
\addlinespace
\multicolumn{7}{@{}l}{\emph{1-h / 24-h}} \\
Independent & 12.23 & 250.5 & 0.66 & 0.88 & 1.002 & 3.3 \\
Gaussian copula & 11.78 & 163.0 & 0.16 & 0.42 & 0.935 & 5.2 \\
Residual bootstrap & 12.16 & 159.7 & 0.17 & 0.12 & 1.026 & \textbf{7.6} \\
Schaake shuffle & \textbf{11.75} & \textbf{156.6} & \textbf{0.09} & \textbf{0.09} & \textbf{0.896} & \textbf{7.6} \\
\bottomrule
\end{tabular}
  \\[2pt]
  \parbox{\columnwidth}{\scriptsize Bold: best value in each task; for shortfall, the value
  closest to the 10\% target.}
\end{table}

\subsection{Effect on Reserve Sizing}
The distortion carries into the reserve held over the horizon. With independent scenarios,
the simulated trajectories contain larger excursions than the observed load, so the estimated
quantile of $D_t$ is too high. At $c_R/c_S=0.1$, the reserve is 9.43~kW at the 24-h horizon and the realized
shortfall rate is 3.3\% against a 10\% target. The Schaake shuffle sets a reserve of 7.59~kW
with a shortfall rate of 7.6\%, and its mean cost is 0.896 against 1.002. Across the three tasks
the cost falls by 7.4\%, 9.1\% and 10.6\%, and the shortfall rate is 10.5\%, 9.3\% and 7.6\%.
The Schaake shuffle also has the lower cost at $c_R/c_S=0.05$ and $0.2$ in every task, and at
$0.2$ the reduction reaches 15.4--17.8\%.

The residual bootstrap shows that a good Stage-3 score does not guarantee a low Stage-4 cost. At 5-min resolution it
matches ramps as well as the Schaake shuffle (KS 0.07) yet costs 0.390 against 0.326, because its
reserve barely changes with the forecast spread at each origin. In this benchmark, the lowest reserve
cost requires both the temporal structure of past errors and the conditional spread of the
forecaster.

\section{Conclusion}

This paper built an open benchmark from measured GPU telemetry of a 448-GPU AI cluster and
evaluated probabilistic forecasts through four stages that end in reserve sizing. The
stage-wise evaluation locates where forecast value is lost. After the mid-May rise in utilization, a quantile
LightGBM model covered 63.4\% of outcomes with a nominal 90\% interval at the 24-h horizon. A
causal online conformal correction restored 87.3\%, and a quantile TCN was close to nominal
without it. At the next stage, calibrated quantiles sampled step by step distorted ramps and
peaks and over-procured reserve held over the horizon. Reordering the samples with past error
trajectories kept the conditional spread, matched ramp and peak statistics, and lowered the
reserve cost by 7.4--10.6\%. These results underscore the need for load scenarios that preserve both forecast-conditioned uncertainty and temporal dependence when sizing reserves over a forecast horizon.

\end{document}